\documentclass[prl,aps,twocolumn,showpacs,amsmath,amssymb]{revtex4}

\usepackage{graphicx}
\usepackage{bm}

\newcommand{\bra}[1]{\langle #1|}
\newcommand{\ket}[1]{|#1\rangle}

\newcommand{\average}[1]{\langle #1 \rangle}

\begin{document}

\title{Orbital contribution to the magnetic properties of nanowires: Is the orbital polarization ansatz justified?}

\author{     M.C. Desjonqu\`eres$^*$, C. Barreteau$^*$, G. Aut\`es$^*$, D. Spanjaard$^{\dag}$}

\affiliation{$^*$CEA Saclay, DSM/DRECAM/SPCSI, B\^atiment 462,
F-91191 Gif sur Yvette, France }

\affiliation{$^{\dag}$Laboratoire de Physique des Solides,
             Universit\'e Paris Sud, B\^atiment 510, F-91405 Orsay, France}

\date {\today}

\begin{abstract}
We show that considerable orbital magnetic moments and magneto-crystalline
anisotropy energies are obtained for a Fe monatomic wire described
in a tight-binding  method with intra-atomic electronic interactions
treated in a full Hartree Fock (HF) decoupling scheme.
Even-though the use of the orbital polarization ansatz with
simplified Hamiltonians leads to fairly good results when the
spin magnetization is saturated this is not the case of
unsaturated systems. We conclude that the full HF scheme is necessary
to investigate low dimensional systems.
\end{abstract}

\maketitle


In the bulk of ferromagnetic transition metals it is well known that
the orbital magnetic moment  {\bf L} is
quenched and that the magneto-crystalline anisotropy energy (MAE) is very small
as a result of strong electron delocalization and crystal
field effects compared with those of intra-atomic Coulomb
interactions. 
In nano-objects the dimensionality or coordination is reduced
so that  the influence of these interactions, responsible for Hund's rules
in the free atom, becomes more and more important and both the
spin and orbital magnetic moments increase dramatically.
This is seen in
experiments on chains of Co atoms at step edges of Pt(997)\cite{Gambardella02} 
and Co single atoms or nanoparticles
deposited on Pt(111) in which orbital moments as large as
1.1$\mu_B$ per atom have been measured\cite{Gambardella03},
associated with a considerable enhancement of the MAE.

On the theoretical side, in the Local Spin Density Approximation
(LSDA) or in simplified tight-binding (TB) Hartree-Fock (HF)
schemes the intra-atomic Coulomb interactions are treated in an
average manner so that the distribution of electrons between the
orbital states of opposite magnetic quantum numbers $m$ is poorly
described, especially in low dimensional systems. As a result these approximations yield
underestimated values of {\bf L}, even though these values
increase when the dimensionality is lowered. Eriksson {\sl et al.}\cite{Eriksson90}
have proposed to correct for this effect by
adding a term proportional to $-\widehat{{\bf L}}^2/2$ in the
Hamiltonian, treated in mean-field, which will be referred to as
Orbital Polarization Ansatz (OPA) in the following. The effect of
this term is obviously to increase $<\widehat{{\bf L}}>$. A more
rigorous way of obtaining both the spin and orbital moments is to
solve the HF equations by taking into account all intra-atomic
terms in the decoupling with all matrix elements of the Coulomb
interaction
$U_{\gamma_1\gamma_2\gamma_3\gamma_4}=\bra{\gamma_1(\bm{r}),\gamma_2(\bm{r'})}\frac{e^2}{|\bm{r-r'}|}\ket{\gamma_3(\bm{r}),\gamma_4(\bm{r'})}$,
where $\gamma_i$ are atomic orbitals, expressed in terms of the
three Racah parameters $A, B$ and $C$, for $d$ electrons\cite{Griffith61}
 and a system of homonuclear atoms. Starting from
this Hamiltonian Solovyev {\sl et al.}\cite{Solovyev98} have
shown, in an elegant work, that the OPA cannot be derived
analytically from the HF Hamiltonian except in some very special
cases and that, even in the latter, the proportionality factor is
not $B$ as usually assumed but $3B/2$. Very recently Nicolas {\sl
et al.}\cite{Nicolas06} have discussed the effect of orbital
polarization, using either a Stoner-like TB Hamiltonian with the
OPA or an HF Hamiltonian in which the one and two orbital matrix
elements of the Coulomb interaction are treated exactly in the
spherical harmonics (SH) basis but three and four orbital terms
are neglected. These latter terms depend both on $B$ and $C$ in the
SH basis which results in a symmetry breaking that they claim to overcome
by averaging over different orbital basis. On the opposite, a recent
work by Xiangang Wan {\sl et al.}\cite{Xiangang_Wan04} is based on
a complete HF decoupling. However their effective intra-atomic
potential (see Eq.4 of their work) is the same as in LSDA+U while
the TB part of their total Hamiltonian is not spin polarized. As a
result when the approximations leading to the Stoner model are
carried out in their Eq.4, it does not lead to the correct Stoner
parameter.

It is thus of fundamental importance to investigate the ability
of the full HF scheme to predict large {\bf L} and MAE in nano-objects,
and to check whether the OPA can account for these effects.
In this paper we compare, on the simple model of a
monatomic wire, the results given by the full HF decoupling and
two currently used simplified Hamiltonians corrected or not by the
OPA term. We use a TB model in a minimal orthogonal basis set of
$d$ valence orbitals $\ket{i,\gamma,\sigma}= \ket{i,\gamma}
\otimes \ket{\sigma}$, of spin $\sigma$ and orbital $\gamma$
centered at site $i$. In the following $\gamma$ will either denote
cubic harmonics (CH)
($\gamma=\lambda=d_{xy},d_{yz},d_{zx},d_{x^2-y^2},d_{3z^2-r^2} $)
or spherical harmonics ($\gamma=m=-2,-1,0,1,2$). Our
Hamiltonian $H$ can be expressed as the sum of a standard one-body
TB Hamiltonian $H_0$ (determined by the bare $d$ level
$\varepsilon_0$, and hopping integrals) and an electron-electron
interaction Hamiltonian $H_{\text{int}}$ in which only on-site
electron-electron interactions are considered. The standard
Hartree Fock decoupling leads to the one-electron Hamiltonian
(denoted as HF1) which, in the second quantization formalism can
be written

 \begin{eqnarray}
 H_{\text{int}}^{\text{HF1}}=& \displaystyle
             \sum_{ \substack{i \gamma_1 \gamma_2 \gamma_3 \gamma_4 \\ \sigma \sigma'} }
             \Big( U_{\gamma_4\gamma_2\gamma_3\gamma_1} \average{c_{i\gamma_4\sigma}^{\dag}c_{i \gamma_3\sigma}  } c_{i \gamma_2\sigma'}^{\dag}  c_{i \gamma_1\sigma'} \nonumber    \\
            &  -     U_{\gamma_4\gamma_2\gamma_1\gamma_3}  \average{c_{i \gamma_4\sigma}^{\dag}c_{i \gamma_3\sigma'}} c_{i \gamma_2\sigma'}^{\dag}   c_{i \gamma_1\sigma} \Big)
\end{eqnarray}

\noindent
 The expressions  of the matrix elements $U_{\gamma_1\gamma_2\gamma_3\gamma_4}$ obviously depend
on the atomic basis, but the resolution of the full Hartree-Fock
Hamiltonian (namely without any approximation) must lead to the
same results whatever the basis. However, the use of CH
is quite attractive for discussing the OPA since  in
this basis the three and four orbital matrix elements of the
electron-electron interaction are proportional to the Racah
parameter $B$ only\cite{Griffith61}. Moreover in CH the different
values of the two orbital matrix elements
$U_{\lambda\mu\lambda\mu}$ and  $U_{\lambda\mu\mu\lambda}$
($\lambda \ne \mu$) only differ by terms proportional to $B$. The
average values: $(1/4)\sum_{\mu,\mu\ne\lambda} U_{\lambda \mu
\lambda \mu}$ and $(1/4)\sum_{\mu,\mu\ne\lambda} U_{\lambda \mu
\mu \lambda}$ are independent of $\lambda$ and are given by $U=A-B+C$
and $J=5B/2+C$\cite{Anisimov_convention} while the one orbital
terms $U_{\lambda\lambda\lambda\lambda}$ are all equal to $U+2J$.
This leads us to define $(U,J,B)$ as a new set of parameters.
 The two orbital terms $U_{\lambda\mu\lambda\mu}$
(resp. $U_{\lambda\mu\mu\lambda}$) can then be expressed in terms
of $U$ and $B$ (resp. $J$ and $B$) while the three and four
orbital terms are proportional to $B$ only. As already stated,
this is no longer true in the SH basis.

When $B$ is neglected in the above Hamiltonian HF1, we recover the
model (hereafter referred to as HF2) that has been used in our
previous studies\cite{Barreteau2001} ($U_{\lambda\mu\lambda\mu}=U$ and
$U_{\lambda\mu\mu\lambda}=J$ for any pair of different orbitals
$\lambda$ and $\mu$ and no three and four orbital terms) in which
spin-flip terms were omitted since the spin-orbit coupling
interaction was not taken into account. Starting from this
Hamiltonian, keeping only the diagonal terms and replacing each
orbital population of a given spin by its average value, leads to
a Stoner-like Hamiltonian (called HF3) that we have also
investigated since it has widely been used in the literature\cite{Ricardo06}:

\begin{equation}
H_{\text{int}}^{\text{HF3}}=\sum_{i \lambda,\sigma} (U_{\text{eff}} N_i -\sigma IM_i/2 ) c_{i\lambda \sigma}^{\dag}  c_{i\lambda\sigma}.
\end{equation}

\noindent In this hamiltonian $I=(U+6J)/5$ is the Stoner parameter
while $N_i$ and $M_i$ are, respectively, the total charge
and moment on site $i$. $U_{\text{eff}}$  is equal to $(9U-2J)/10$ if
one derives HF3 from HF2 as explained above. Since here we are interested
in systems with geometrically equivalent atoms (i.e., $N_i=N, M_i=M$) we can choose
the energy zero in all hamiltonians as $\varepsilon_0 + U_{\text{eff}}N$
so that the first term in Eq.(3) disappears from the total hamiltonian HF3.

The spin magnetism is governed by the Stoner parameter $I$ that
will be kept constant in all our calculations and determined so
that it reproduces the experimental value in the bulk bcc phase.

From the above discussion it is clear that HF2 differs from HF1 by
terms proportional to $B$, this is also true for HF3 as far as
this hamiltonian is justified. Eriksson {\sl et
al.}\cite{Eriksson90} have proposed to introduce an OPA term to
account for this difference. This term is written in meanfield $\Delta
E_{\text{OP}}=-\frac{1}{2}B \sum_i \average{\bm{L}_i}^2$ which
reduces to
 $-\frac{1}{2}B \sum_i \average{L_{iz}}^2$ when the spin and orbital moments have
the same quantization axis $z$ (which is strictly verified along high symmetry directions).
 The corresponding Hamiltonian is then:

\begin{equation}
H_{\text{OP}}=-B \average{L_z} \sum_{i \gamma \gamma'}  [L_z]_{\gamma \gamma'} c_{i \gamma \sigma}^{\dag} c_{i \gamma' \sigma}
\end{equation}

 \noindent
where $[L_z]_{\gamma \gamma'}$ are the matrix elements of the
local orbital moment operator $L_{iz}$. $[L_z]_{\gamma \gamma'}$
is spin independent and  diagonal in the SH basis when the orbital
momentum quantization axis of the SH orbitals is rotated so that
it coincides with the spin quantization axis. This is no longer
true if the SH orbital momentum axis is along a crystallographic
axis which is not parallel to the spin quantization axis, or when
$[L_z]_{\gamma \gamma'}$ is expressed in the CH basis. Finally the
last term of our Hamiltonian takes into account the intra-atomic
spin-orbit interactions determined by the spin-orbit coupling
parameter $\xi$.

A monatomic wire of a transition metal is a handy system to
compare the results given by the various models described above.
The parameters of the
model are chosen to mimic Fe which is assumed to have $N=7$ valence
{\it d} electrons per atom in the bulk as well as in the wire. The
hopping integrals $dd\sigma$, $dd\pi$ and $dd\delta$ are chosen
proportional to (-6,4,-1) and decrease with the interatomic
distance according to a $R^{-5}$ law. The numerical value of
$dd\sigma$ is fitted to the bulk $d$ band width of Fe ($W_d=6$eV)
which leads to $dd\sigma=-0.749$eV at the bulk nearest neighbor
distance (d=4.7a.u.). First and second nearest neighbor hopping
integrals have been taken into account. The Stoner parameter
is $I=0.67$eV. The
spin-orbit coupling parameter is taken from a previous work
($\xi=0.06$eV)\cite{Autes06}. It is well known that the parameter
$U$ is strongly screened in metals. In particular in a
recent paper Solovyev\cite{Solovyev05} has shown that this
parameter is almost independent of the bare interaction. From
Fig.1 of this reference it can be deduced that $U\simeq J$ in Fe\cite{Solovyev_convention}.
In that case $I=7J/5$ so that $U=J=0.48$eV, a numerical value in
good agreement with that given by Solovyev. Finally, as in
previous works\cite{Solovyev05}, we have taken $B=0.14J$\cite{Anisimov_convention}.

When applied to bulk Fe, the complete HF decoupling
yields $\average{2S_z}=2.12\mu_B$ and $\average{L_z}=0.08\mu_B$ when $B=0$ (HF2 model), and
$\average{2S_z}=2.11\mu_B$ and $\average{L_z}=0.12\mu_B$ when $B$ is taken into account (HF1 model).
Then, we have compared the results derived from the five models
(HF1, HF2 and HF3 with and without $H_{OP}$)
for the spin and orbital moments with magnetizations
along the wire ($\theta=0$) and perpendicular to it
($\theta=\pi/2$) and the corresponding magnetocrystalline anisotropy energy
(MAE) $\Delta E=E_{tot}(\theta=\pi/2)-E_{tot}(\theta=0)$ where $E_{tot}$
is the total energy per atom of the system. Two interatomic
distances have been considered: the bulk interatomic distance at
which the spin magnetization is saturated and a shorter distance
(4.25a.u.) corresponding to unsaturated spin moments. The results
are given in Table 1.

\begin{table}
\begin{center}
\begin{tabular}{|c|c|c|c|c|c|}\hline
& HF1 & HF2 & HF2 & HF3 & HF3 \\
\hline
                          &       &      & OPA &      & OPA \\
\hline
                    \multicolumn{6}{|c|}{$d=4.7$a.u.} \\
\hline
$\average{2S_z(0)}$       & 3    & 3    & 3    & 3    & 3     \\
$\average{2S_z(\pi/2)}$   & 3    & 3    & 3    & 3    & 3     \\
$\average{L_z(0)}$        & 1.45 & 0.37 & 1.31 & 0.37 & 1.31  \\
$\average{L_z(\pi/2)}$    & 0.49 & 0.25 & 0.61 & 0.25 & 0.60  \\
MAE                       & 23.4 & 0.7  & 22.3 & 0.6  & 22.3  \\
\hline
                  \multicolumn{6}{|c|}{$d=4.25$a.u.} \\
\hline
$\average{2S_z(0)}$       & 1.51 & 1.24 & 1.23 & 0.94 & 0.78  \\
$\average{2S_z(\pi/2)}$   & 1.51 & 1.23 & 1.24 & 0.93 & 0.94 \\
$\average{L_z(0)}$        & 0.33 & 0.19 & 0.39 & 0.24 & 1.07  \\
$\average{L_z(\pi/2)}$    & 0.21 & 0.10 & 0.18 & 0.08 & 0.15  \\
MAE                       & -0.7 & -0.3 & 1.5 & 0.0 & 6.2  \\
\hline
\end{tabular}
\newline
\caption{The spin ($\average{2S_z}$) and orbital ($\average{L_z}$) magnetic moments (in
$\mu_B$ per atom) for a monatomic Fe wire and two magnetization orientations
(parallel ($\theta=0$) and perpendicular ($\theta=\pi/2$) to the wire) and the
corresponding magnetocrystalline anisotropy MAE ($E_{tot}(\pi/2)-E_{tot}(0))$ in
meV per atom for two interatomic distances.}
\end{center}
\end{table}

Let us first discuss the wire at the bulk interatomic distance.
All models agree to predict saturated spin magnetization, i.e.,
the spin magnetic moment is $3\mu_B$ to less than a few
$10^{-3}\mu_B$. As a consequence the effective atomic orbital
levels with down spin are identical in HF2 and HF3 models since $U=J$. This is no longer
true for the up spin orbitals for which the atomic levels are orbital
dependent with HF2 and not with HF3. However the average atomic
level is the same in both models. Therefore the orbital moment,
which arises only from the spin down band, the spin up band being
filled, is almost identical in both models similarly to the total
energy (see Table 1). As expected the orbital moments for both magnetization
orientations and the associated MAE, even though reinforced
compared to the bulk ones, are largely underestimated by the HF2
and HF3 models with $B=0$ compared to those predicted by the
complete HF decoupling (HF1). When the OPA term is added to the HF2 and
HF3 hamiltonians, the results given by the latter models become in
fair agreement with those obtained from HF1 for the orbital
moment while the MAE is well reproduced.

The above trends completely change when the interatomic distance
is shortened to 4.25a.u.. It is first seen that the spin moment
depends on the model. In this respect the HF2 model is much better
than the HF3 one. Moreover, taking into account the OPA term leads
to an increase of the orbital moments for both magnetization
orientations which are rather close to the HF1 results for HF2 but
not for HF3.

\begin{figure}[!fht]
\begin{center}
\includegraphics*[scale=0.4,angle=0]{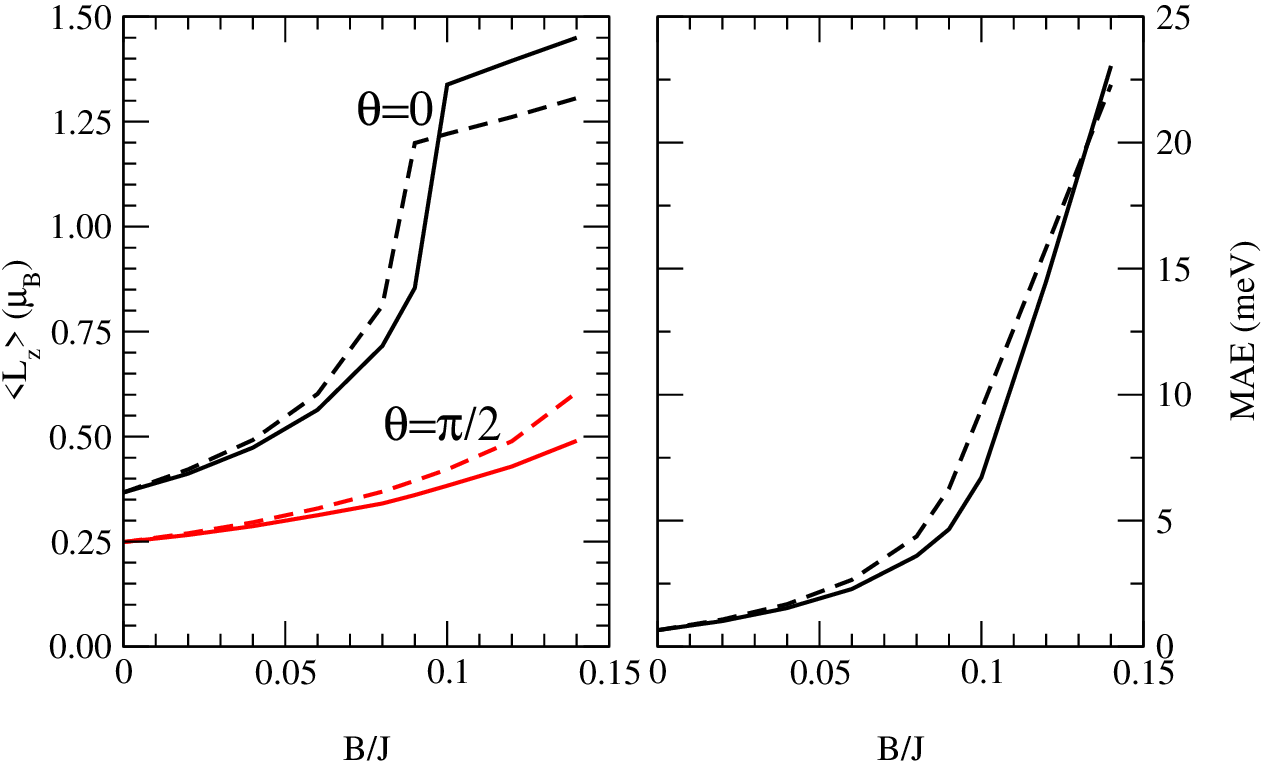}
\end{center}
\caption{$\average{L_z}$ and MAE as a function of $B/J$ from HF1
and HF2 (HF3 results are undistinguishable from the HF2 ones) for
a magnetic Fe monatomic wire ($d=$4.7a.u.).} \label{fig:B_sur_J}
\end{figure}

To summarize this discussion we can state that the OPA is rather
good for saturated spin magnetization while for the unsaturated
case it leads to results depending critically on the
approximations made concerning the electron-electron interaction
hamiltonian. In order to verify that the good performance of the
OPA for the saturated spin magnetization is not due to the
particular value of $B$, we have studied the variation of $\average{L_z}$ at
$\theta=0$ and $\theta=\pi/2$ and the associated MAE as a function
of the ratio $B/J$. The results (Fig.\ref{fig:B_sur_J}) show that the OPA gives the
right trends on the full domain of $B/J$ values that we have
investigated. In particular an abrupt variation of $\average{L_z}$ at
$\theta=0$ occurs around a critical value of $B/J\simeq 0.09$ above
which the upper $\delta$ band (the corresponding eigenfunctions
being mostly linear combinations of SH with $|m|=2$) of minority spin becomes empty.

\begin{figure}[!fht]
\begin{center}
\includegraphics*[scale=0.45,angle=0]{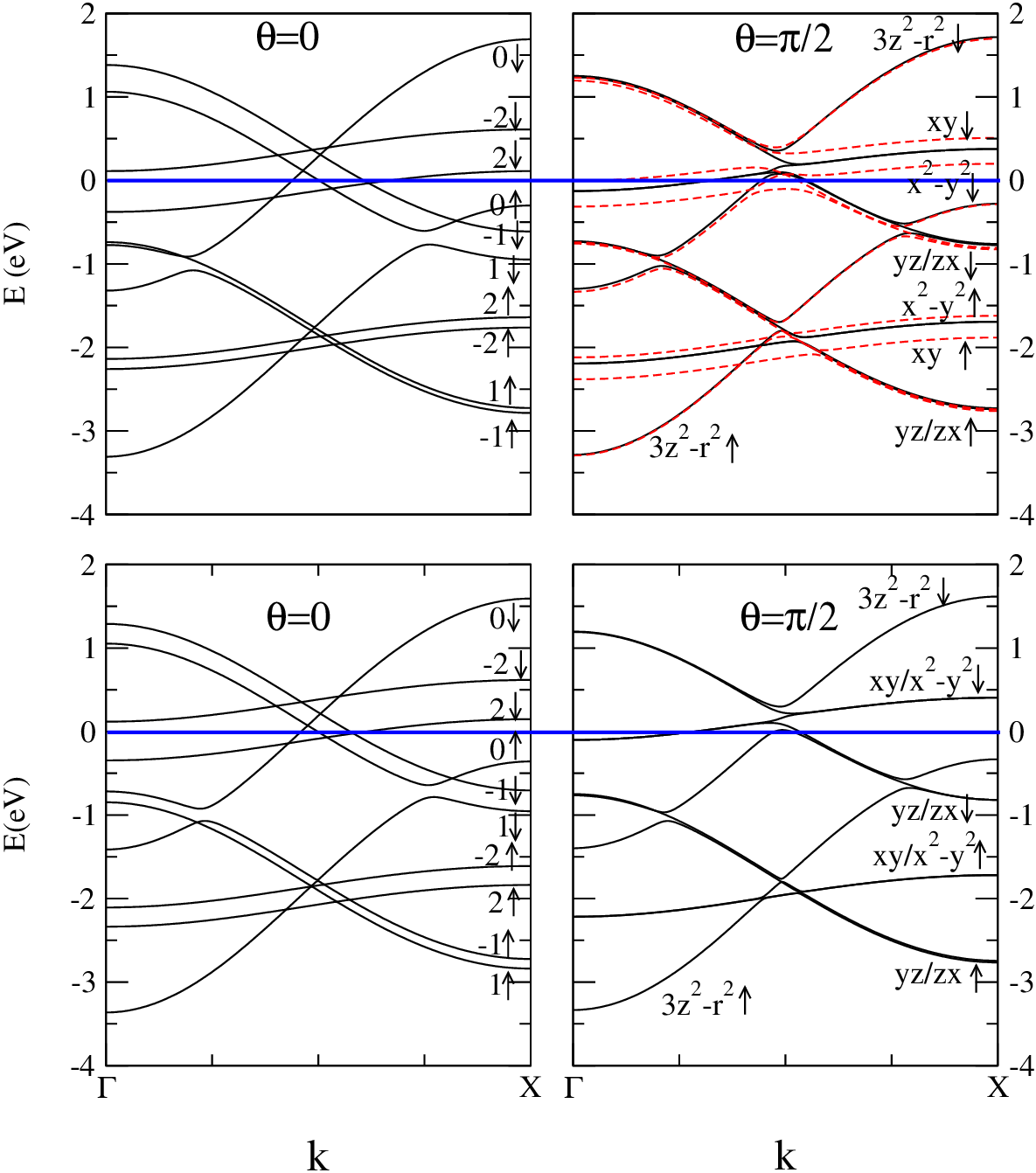}
\end{center}
\caption{HF1 (top) and HF2 (bottom) band structure (referred to
the Fermi level) for a magnetic
Fe monatomic wire ($d=$4.7au) with a magnetization parallel
($\theta=0$) and perpendicular ($\theta=\pi/2$) to the wire. All
results are obtained for $U/J=1$ and $B=0.14J$  save for the
dotted band structure of the the top right panel obtained for
$U/J=1.34$ and $B=0.14J$.} \label{fig:bandstructure}
\end{figure}

Even if the OPA works reasonably in the saturated spin
magnetization case for determining $\average{L_z}$ and the MAE,
this does not mean that it reproduces the band structure
correctly. Let us first note that for $\theta=0$, the
eigenfunctions have a largely dominating single SH character while
at $\theta=\pi/2$ they are almost pure single CH orbitals. The
band structures corresponding to HF1 and HF2  are drawn in
Fig.\ref{fig:bandstructure} (the band structure of HF3 is close to
that of HF2). At first sight they look quite similar. However a
closer examination reveals some differences. Let us first comment
on the majority spin bands at $\theta=0$. While the splittings of
the $|m|=2$ ($\delta$) and $|m|=1$ ($\pi$) bands are respectively
given by $2\xi$ and $\xi$ with the HF1 model, they become
$2\xi-4B\average{L_z}$ and $\xi-2B\average{L_z}$ with both the HF2
and HF3 models, respectively. In addition the $m$ character of the
bands is reversed, i.e., the $m$=2(1) band is above the $m$=-2(-1)
band in the HF1 while it is the opposite with the HF2 and HF3
models. This inversion does not occur in the minority spin bands
and the splittings of the $\delta$ and $\pi$ bands are not exactly
the same with the HF2 and HF3 as with the HF1 models. At
$\theta=\pi/2$ all models agree that for $U=J$ there are almost no
band splittings and that the removals of degeneracy around the
midpoint between $\Gamma$ and $X$ are more pronounced in the
minority bands than in the majority ones.

\begin{figure}[!fht]
\begin{center}
\includegraphics*[scale=0.4,angle=0]{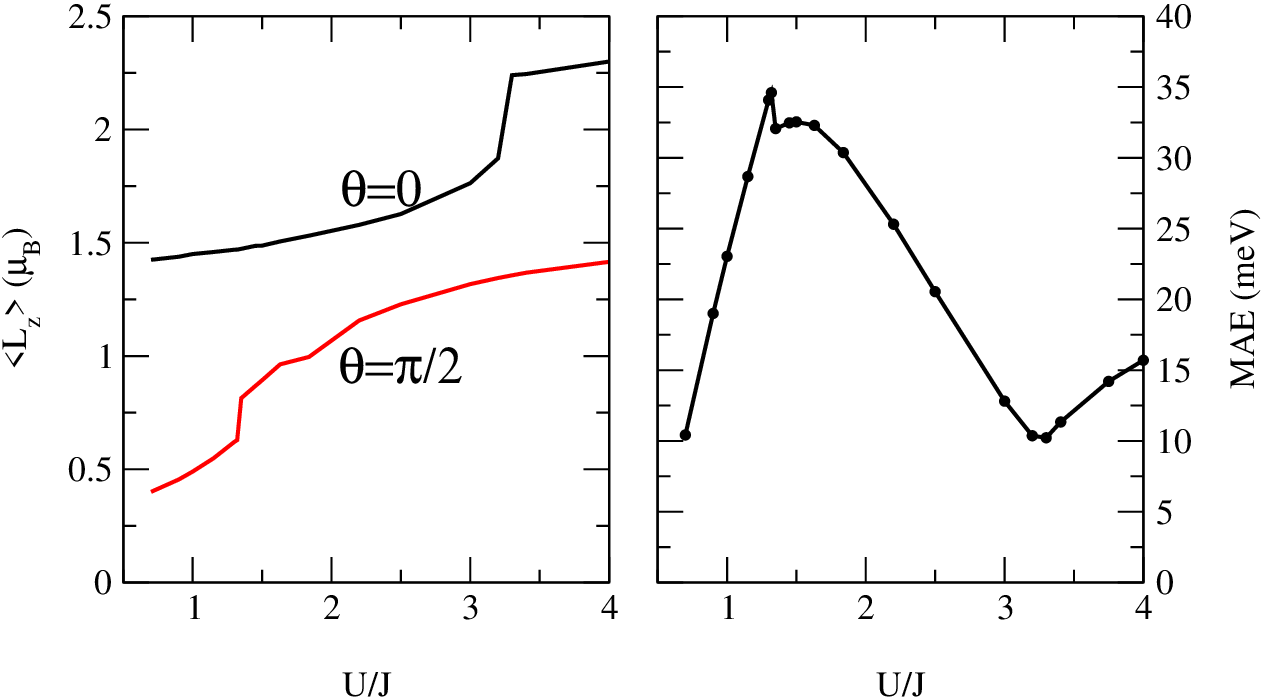}
\end{center}
\caption{$\average{L_z}$ and MAE as a function of $U/J$ from HF1
for a magnetic Fe monatomic wire ($d=$4.7au).} \label{fig:U_sur_J}
\end{figure}

Finally it is interesting to study the variation of $\average{L_z}$ and of
the MAE with the HF1 model when the ratio $U/J$ is varied by keeping
the Stoner parameter fixed. Indeed this ratio is not perfectly
known. The results are shown in Fig.\ref{fig:U_sur_J}. Abrupt variations of $\average{L_z}$
are observed at $U/J\simeq 1.34$ when $\theta=\pi/2$ and
$U/J\simeq 3.25$ when $\theta=0$. They correspond respectively to the occurrence
of a splitting of the $\delta$ bands (see
Fig.\ref{fig:bandstructure}) and to the complete filling of the lowest $\delta$ band of
minority spin. These abrupt changes of $\average{L_z}$ are
associated with a change of sign of the variation of the MAE as a
function of $U/J$.

In conclusion we have studied orbital polarization effects for a
Fe monatomic wire with various HF Hamiltonians in a tight-binding
scheme: a full HF Hamiltonian (HF1) including all the Coulomb
interaction matrix elements, a simplified one (HF2) neglecting the
Racah parameter B, and finally a Stoner-like Hamiltonian (HF3).
OPA has then been reintroduced in HF2 and HF3 as proposed by
Eriksson {\sl et al}\cite{Eriksson90}. 
With HF1 we predict that very large values of {\bf L} and MAE are possible
in agreement with existing experiments.
The same trends are obtained by adding the OPA to
simplified Hamiltonians when the spin moment is saturated,
however noticeable differences appear in the band structure since
some splitting and band characters are wrongly reproduced. 
This fair agreement strongly deteriorates
when dealing with an unsaturated system, especially with the
Stoner-like model. It is thus of prime importance to use the HF1 model
for the study of systems with much more complex
geometries (surfaces, clusters, break junctions), in a
realistic $s$, $p$ and $d$ basis set, or to implement it in
ab-initio codes. Indeed from our results giant anisotropy of
magneto-resistance in low dimensional systems such as
break junctions is expected\cite{Viret2006}.


\begin{thebibliography}{}
\bibitem{Solovyev_convention}
Note that in Solovyev work $U$ denotes
$U_{\lambda\lambda\lambda\lambda}$ and thus corresponds to $U+2J$
with our definition of $U$ and $J$.
\end{thebibliography}
\end{document}